# MOBILE CODES LOCALIZATION IN AD HOC NETWORKS: A COMPARATIVE STUDY OF CENTRALIZED AND DISTRIBUTED APPROACHES

Youcef Zafoune[1], Aicha Mokhtari[1] and Rushed kanawati[2]

[1]Department of Computer Science, USTHB University, Algiers, Algeria
yzafoune@usthb.dz , amokhtari@usthb.dz
[2] Institute Galilée, Paris13 University, Paris, French
rushed.kanawati@lipn.univ-paris13.fr

## ABSTRACT

*This paper presents a new approach in the management of mobile ad hoc networks. Our alternative, based on mobile agent technology, allows the design of mobile centralized server in ad hoc network, where it is not obvious to think about a centralized management, due to the absence of any administration or fixed infrastructure in these networks. The aim of this centralized approach is to provide permanent availability of services in ad hoc networks which are characterized by a distributed management. In order to evaluate the performance of the proposed approach, we apply it to solve the problem of mobile code localization in ad hoc networks. A comparative study, based upon a simulation, of centralized and distributed localization protocols in terms of messages number exchanged and response time shows that the centralized approach in a distributed form is more interesting than a totally centralized approach.*

## 1. INTRODUCTION

Mobile ad hoc networks is an emerging technology that has gained a lot of interest due to the wide spreading of mobile and wireless devices (i.e. personal assistance, net-books, embedded computers, etc.). An ad hoc network can be defined as a collection of mobile entities interconnected by a wireless technology, forming a temporary network without the help of any administration or any fixed support. This class, which tries to extend the mobility notions to all the composites of the environment, is characterized by a dynamic topology, limited bandwidth and energy power. The absence of the fixed infrastructure obliges the mobile units to behave as routers which participate in the discovery and the maintenance of the paths for the network hosts. These mobile hosts themselves which form, in an ad hoc manner, a way of global architecture which can be used as an infrastructure of the network. Consequently, any management in these networks is a distributed type. The distributed applications call most often at all nodes in the network, resulting in a response time and a flood of messages quite important. Is it possible to envisage a centralized management as an alternative? That is the aim of our research in this paper.

Our alternative consists in designing mobile centralized server in ad hoc network. The aim of this centralized approach is to provide permanent availability of services in ad hoc networks. Our proposed approach for ad hoc networks where it is not obvious to think about a centralized management is based on mobile agent technology. The mobile code technology, issued from the domain of artificial intelligence under the name of mobile agent, is now growing fast. This is promising in many fields of application, especially in the field of nomadic computer science and in the distributed applications in large scale networks (such as information retrieval and electronic commerce in the web). A mobile agent is a software agent that has the ability to move from one hosting node to another without interpreting its execution. The challenge with using the mobile agent technology is to endow the mobile server with the capacity to select a hosting node to enhance the obtained service.

In order to evaluate the performance of the proposed approach, we apply it to solve the problem of mobile code localization in ad hoc networks. The localization mechanism has become a major component of future services with the proliferation of communicating objects. Many localization methods of mobile objects in different kind of networks have emerged. In our study, we are interested in the problem of the mobile code localization in mobile ad hoc networks. While in the network with fixed stations, the mechanisms of localization of mobile code are classic, it is not the same for mobile ad hoc networks characterized by a dynamic network topology.

In the second section of this paper, we present the problematic of mobile object localization. In the third section, we describe our proposed mobile code's localization protocols in the ad hoc network. Then, a comparative study of centralized and distributed approaches, based upon a simulation, is presented in the fourth section. Section five presents an improvement of the centralized protocol with the study of the performance. Finally, we present our conclusions and future work in the last section.

## 2. PROBLEMATIC OF MOBILE OBJECT LOCALIZATION

Any mobility mechanism must ensure that any communication posterior to a migration will not be impaired by it, namely that two objects should still be able to communicate even if one of them has migrated. Such mechanism is referred to as an object localization mechanism (agent or mobile station), which enables the communication between these objects in spite of the change of their positions.

The localization mechanism is used generally in the first phase of information routing, more exactly in the routes discovery procedure which allows any object in a network to dynamically discover a path toward any other object of the network. The localization management puts forward two basic operations: the research and updating. The research operation allows localizing an object when this localization is needed. The updating operation allows to inform the network node mobile object new position when this latter moves within the network. Thus, it is important to establish:

- When to update the localization information? In reactive or proactive manner?

- Which nodes should take place information of localization and where to send the updating? It could be some specific nodes (centralized architecture) or all the networks nodes (distributed architecture). Furthermore, each of these nodes can maintain the localization information of some specific nodes (hierarchical structure) or of all the network nodes (non hierarchical structure).

Different contexts have been presented in literature, in which the localization problem of mobile objects (station or agent) in networks occurs. From the onset of cellular mobile networks emergence, this problem has been studied for the location of mobile stations. The existence of fixed infrastructure (basic stations) in these networks has facilitated the task to develop robust solutions to this problem. This was not the case with the introduction of mobile ad hoc networks, due to the absence of any fixed administration. Different tracking nomad stations services, based on different criteria, were then proposed (see Section 2.3). The development of mobile agent technology with its first use in networks with fixed stations raised the problem of the localization of mobile code, where simple protocols have been proposed (see Section 2.1). The use of mobile agent in mobile ad hoc networks makes the problem of localization of mobile code more complex, on the one hand, because of the lack of fixed infrastructure in these networks as mentioned previously, and on the other hand, because of double mobility management, namely, the code resorts across the network, as well as stations in the network.

## 2.1. Mobile Code Localization in Fixed Network

The mobile object describes here an agent or mobile code which moves in a network with fixed stations. Two mechanisms of localization in this case are principally used [1]:

- The first mechanism dynamically creates a chain of forwarders in order to localize a mobile code. Forwarding techniques were first introduced in distributed operating systems like DEMOS/MP [22] for finding mobile processes. The mechanism is straightforward: every time a code leaves a station, it leaves a special reference, called forwarder, which points to the next destination, i.e. the mobile code on a reception site. This dynamically creates a chain of forwarders which allow to localize the mobile code; when a forwarder receives a message, it sends it to the next destination (it is possible that is the mobile code).

The description of the protocol is as follows:

1- Upon migration, a mobile code leaves a forwarder on the current site;
2- This forwarder is linked to the mobile code on the remote host;
3- No communication can occur when the code is migrating;
4- When receiving a message, the forwarder sends it to the next hop (possibly the mobile code);
5- Any successful communication places the mobile code one hop away of the caller.

- An alternative to the forwarders approach for locating a mobile code is to use a centralized server to accomplish this task.

Servers like this are widely used in the Internet. For instance, the Domain Name Server [19] uses hierarchically organized servers to associate location (IP address) to symbolic name.
The idea behind location server is simple: the server keeps a trace of the mobile code's position in a data base. Each time a mobile code migrates, it informs the server of its new position. When the source site wants to reach its mobile code, it sends a message to the last known position of the mobile code; if this communication fails, then the source sends a localization request to the server. The description of the protocol used by the source and the mobile code to communicate with the server is as follows:

.The Mobile Code:
1- Performs the migration;
2- Sends its new location to the server.

.The Source:
1- Issues a message to the mobile code with the recorded location.
   Upon failure goes to Step 2;
2- Queries the server to have the current location of the mobile code;
3- Issues a message to the mobile code with the location provided by the server.
   Upon failure, returns to Step 2.

## 2.2. Mobile code localization in ad hoc mobile network

The localization problem of mobile code in ad hoc networks becomes more complex, since it is to localize an object of double mobility, first, in relation to the mobility of its physical support in a dynamic environment, the station, and that of the code through the network stations.
In order to solve this problem, this double mobility makes us think to combine one of the mobile code localization technical in fixed networks with the mobile stations localization services in ad hoc network (section 2.3). We notice that the working of the first distributed protocol, based on the chain of forwarders, in the mobile network poses the problem of the breaking of the chain, caused by the mobility of the marked stations. Whereas the second centralized protocol based on a server, appears to be not adapted to an ad hoc mobile network, due to the absence of a fixed infrastructure and the permanent stations mobility in the network.
In section 3 we present our proposed localization protocols of mobile code in mobile ad hoc network.

### 2.3. Mobile station localization in an ad hoc network

The mobile station localization mechanism is used in several position-based routing protocols that have been developed for mobile ad hoc networks. Many of these protocols assume a location service is available which provides location information on the stations in the network. A location service, which tracks the position of mobile stations in the network and route messages between any two stations, allows to reduce overhead of route discovery since the establishment and maintenance of routes is (usually) not required in a protocol that uses location information for routing. There are two general types of location services: proactive location services and reactive location services [6]:

- Proactive location services are those protocols that have station exchange location information periodically.
- Reactive location services query location information on an as needed basis.

The following are some of these technical:

- GLS (Grid Location Service) which is proactive technical based on a decomposition of network geographical space in predefined grids [16].

- SLS (Simple Location Service) communicates the localization information to solely its neighbors [8].

- LEAP (Legend Exchange and Augmentation Protocol) manipulates global localization table, which when migrating from one node to another in the network, puts together the localization information of each node went through [14].

- DRM (Dead Reckoning Method) is a service based on localization prediction technical [15].

- RLS (Reactive Localization Service) which is, contrary to the preceding technical, of a reactive type based on localization request diffusion first to the neighbors, and in case of failure will be propagated within all the network [8].

- LOTAR (Location Trace Aided Routing protocol) based on the research zone reduction by predicting the rote life span [8].

- PLS (Predictive Location Service) is hybrid localization service (proactive and reactive) based on localization information flooding within the network when not available in the requester node localization table [17].

In our implementations, we used each time one of location services of mobile stations depending on the mobile code localization protocols studied.

## 3. LOCALIZATION PROTOCOLS IN AD HOC NETWORK

Before describing our principally approach, of centralized type, we present first our protocols, of distributed type, for the localization mechanism of mobile code in ad hoc network, which have served as a reference for evaluate our centralized approach.
Note that we have used in the two approaches, the RLS service for localizing the nomadic stations in the ad hoc network. The service RLS, of reactive kind and based on localization request diffusion neighbours first, best suited for our protocols.

### 3.1  Distributed protocol

The distributed localization protocol of mobile code in mobile ad hoc networks is an adaptation of the distributed protocol in the fixed networks described above in section 2.1. Because of the dynamic topology of ad hoc networks, the mobile code localization in a distributed protocol must go through the reestablishment of the chain of forwarders when it is broken. We use two solutions to this problem of the breaking of the chain:

### 3.1.1 Proactive Protocol

The first solution (proactive type) tries to maintain the chain of forwarders in an established state; to do that, it must every time be established immediately after its breaking:

1- Every time a mobile code migrates, it leaves a forwarder on the current station which becomes a marked station. This forwarder is linked to the mobile code on the next station and this creates dynamically a chain of forwarders

2- The neighbors marked stations exchange periodically and continuously check messages, in order to assure the maintaining of the chain in an established state.

3- When a contact is lost between two neighbors marked stations, from $i$ to $j$, the station $i$ emits diffusion for the localization of $j$ in order to re-establish the contact with it. The chain is thus re-established by adding new stations between $i$ and $j$ (figure 1).

4- The marked stations of a higher order which intercept the diffusion request, answer the station $i$ in order to optimize the length of the chain, and this in the case where there is such a station which is nearer from $i$ than from $j$. The chain is re-established and updated (figure 2).

5- The mobile code localization will be will be then take place in the same manner as in the case of a fixed network, since the chain is supposed to be established all the time.

The figure 1.a shows an example of marked stations chain in an ad hoc network, from $a$ to $h$ order. With reference to a relative move of the station $j$, the chain is broken at the level of the station $i$ which emits the localization process of $j$. At the same time, this latter emits the same process for the localization of $k$ since it has lost the contact with its next marked station.

The re-establishment of the chain is shown in the figure 1.b which is done by inserting of new stations in the chain ($m$ and $n$ in the example).

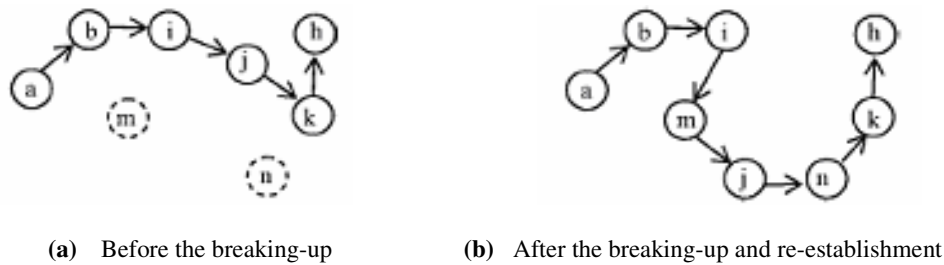

   **(a)**   Before the breaking-up       **(b)**   After the breaking-up and re-establishment

Figure 1.   Chain re-established without optimization

The re-establishment of the chain of the figure 2, always broken at the level of $i$, is achieved in optimal manner since the station $k$ of a higher order than $j$ is within reach of $i$.

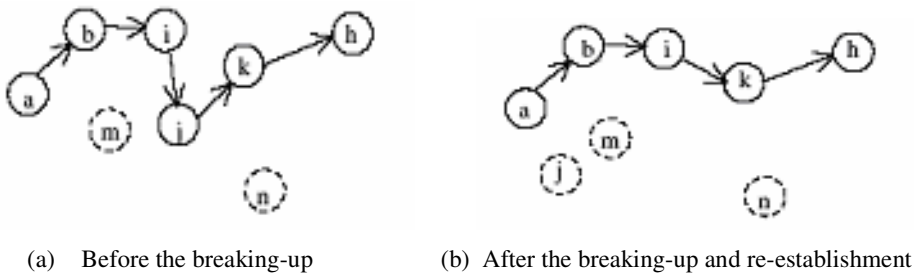

   (a)   Before the breaking-up       (b)   After the breaking-up and re-establishment

Figure 2.   Chain re-established with optimization

### 3.1.2 Reactive Protocol

The second solution (reactive type) is based on the re-establishment of the chain, when it is broken, just when the agent is localized:

1- Same step as in the case of the proactive protocol.
2- To localize a mobile code, the mother station of the code contacts the first marked mobile station in the chain, which in its turn contacts the next marked station, and so on until the mobile code is localized.
3- When a marked mobile station *j* is not accessible by a station *i* (the chain is broken), a limited diffusion is emitted for the search of the station *j* (figure 1).
4- The marked stations of a higher order which intercept the diffusion request, answer the station *i* in order to optimize the length of the chain, and this in the case where there is such a station which is nearer from *i* than from *j* (figure 2).

## 3.2    Centralized protocol

To palliate the fixed infrastructure absence, we propose the use of a virtual server based on mobile agent technology, in charge of the management of the database containing information about the mobile code localization in ad hoc network. The steps of the protocol are as follow:

1- Election of a mobile station as the server, closest to the centre of gravity of the network.
2- A mobile agent is placed on the station elected witch will have as a task the management of the positions database.
3- When the mobile station is away from the centre of the network or for the spent of battery causes, another station is elected for its replacement.
4- The agent placed on the replaced station migrates towards the elected station, where it continues the database management.
5- The mobile code localization will be then take place by sending a request to the mobile server which by consulting the database, gives an answer by sending a message conveying the looked for position, i.e. the station hosting the code.

In figure 3 for example, the station s, which is elected for host the mobile agent, is designed as the server and the stations k, n and m, which are closest to the centre of gravity of the network, are the candidates for replacing s when it is away from the network centre.

Note that this protocol assume that there is no partitioning in the ad hoc network; i.e. all the nodes of the network belong to a same and unique partition.

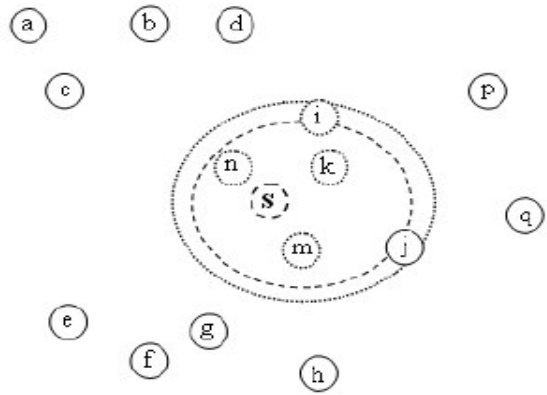

Figure 3. Mobile server in an ad hoc network

The election of the server may be under different selection criteria, such as energy, the degree of connectivity or stations mobility.  We used the proximity to the centre of gravity as a criterion for selection, to allow quick access to a server assumed all the time positioned in the centre of the network. The computing of the gravity centre of a network with n nodes, where $(X_i, Y_i)$ are positions of each node i, is as follow:

$$Xv = 1/n \sum_{t=1}^{n} Xi \qquad\qquad Yv = 1/n \sum_{t=1}^{n} Yi$$

Where $(X_v, Y_{vj}$ define the position of the centre of gravity.

The calculation of the distance of a station $N_i$ from the centre of gravity V, to elect one that is closest to the centre, is as follows:

$$Dist(Ni, V) = \sqrt{(Xv - Xi)^2 + (Yv - Yi)^2}$$

Also note that the positions of a node in the network are provided by the simulator used in our implementation. In practice, it is assumed that uses GPS system or other positioning service.

## 4. EXPERIMENTATION RESULTS

### 4.1 Simulation environment

The performance evaluation of the two localization protocols described earlier has been realized with the Network Simulator NS2 in its version 2.30. The mobility model used in our study is the Random Waypoint Model (RWM) which allows to achieve the simulations according to a random motion of the nodes, where each node has to choose a destination in a random way (belonging to a motion space) and a moving speed (also limited in an interval).

The environment of our simulation is characterized as follows:
- motion area: 1000 x 500 meters$^2$
- number of nodes in the network: 25
- communication range of mobile nodes: 250 meters

The study of the behaviour of the two protocols is done according to the three following parameters:

#### 4.1.1 Mobility of nodes

The mobility of the nodes is defined by their relative motion, which characterizes the way nodes move relatively to each other. The average mobility of a node $i$, $Mi$, is estimated by:

$$Mi = (1/(T - \Delta t)) \sum_{t=0}^{T-\Delta t} |Ai(t + \Delta t) - Ai(t)|$$

Where $T$ is the simulation period, $\Delta t$ is the computation step and $Ai(t)$ represents the average distance separating a node $i$ from all the other nodes at instant $t$ and evaluated in a network with $n$ nodes, as follows:

$$Ai(t) = (1/(n-1)) \sum_{j=1}^{n-1} Dist(Ni, Nj)$$

Then the average mobility of all nodes is defined by:

$$Mob = (1/n) \sum_{i=1}^{n} M_i$$

We distinguish different degrees of mobility: low, medium and high corresponding respectively to the following values of mobility: $0 < Mob <= 3$; $3 < Mob <= 8$ and $8 < Mob$.

In our simulation, we have used three values of mobility for the nodes: 1.5, 5 and 10.

### 4.1.2  Mobility of code

Likewise, we have used the three degrees of mobility for the code: low, medium and high mobility. It corresponds to a number of jumps of the code through the network mobile nodes per time unit.

### 4.1.3  Load of localization request

The load is defined by a poissonian distributed law with parameter $\lambda$. The values taken by this parameter for our simulation are:

0.1, 0.125, 0.167, 0.25, 0.5, 1 and 2 corresponding respectively to period $(1/\lambda)$:

10s, 8s, 6s, 4s, 2s, 1s and 0.5s after which a node generates a new localization request. Thus, the more the localization request load is important, the less is the delay separating two successive requests.

The performance evaluation criteria of the two protocols are:

### 4.1.4  Number of messages

This metric represents the average number of messages required for achieving mobile code localization. It is computed as follows:

Nb_msg = (total number of messages exchanged during the simulation) / (number of localization requests).

### 4.1.5  Response time

This metric represents the average time response for the localization of the mobile code; it is computed as follows:

Rtime = (total time of the all localization requests) / (number of localization requests).

## 4.2  Results and Discussion

The comparative study between the reactive and the proactive distributed protocol has shown [28] that the reactive protocol is by far the less costly and the more adapted in ad hoc mobile networks than the proactive protocol. Since the reactive protocol outperforms the proactive protocol, we present in next paragraph a comparative study between the reactive approach of distributed protocol and the centralized protocol.

### 4.2.1  Distributed (reactive type) vs. centralized protocol

- **Average Messages Number**

Figures 4 and 5 shows respectively the performances of both the distributed and the centralized protocols, in terms of average messages number per localization request, according to the requests load, the mobility of nodes and the mobility of code.

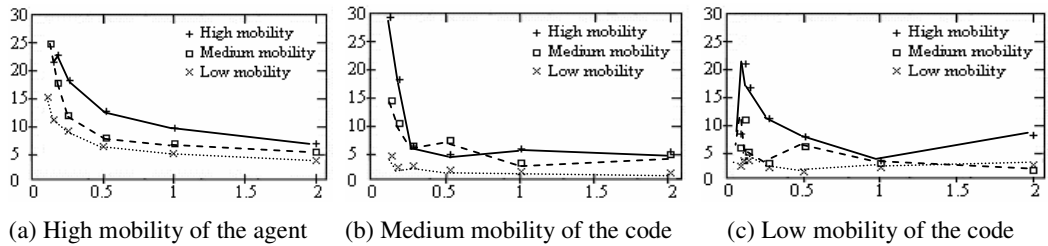

(a) High mobility of the agent    (b) Medium mobility of the code    (c) Low mobility of the code

Figure 4.   Average message number according to requests load (reactive protocol)

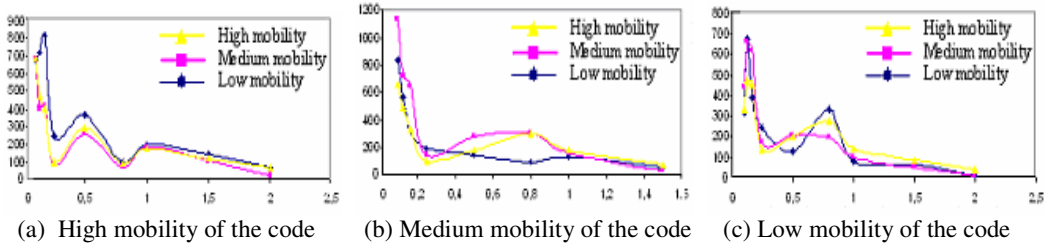

|     |     |     |
| --- | --- | --- |
| (a)  High mobility of the code | (b)  Medium mobility of the code | (c)  Low mobility of the code |

Figure 5.  Average message number according to requests load (centralized protocol)

We remark that the distributed protocol is much less costly in term of the average messages number (30 in maximum) than the centralized protocol (more than 700). This is principally due to the messages of control and of diffusion necessary for the update of the databases of the localization at the server level in proactive way and for the management of the mobile server which is not the case is the distributed approach.

- **Response Time**

Figures 6 and 7 show respectively the performances of the two protocols, distributed and centralized one, in terms of average response time per localization request, according to the requests load, the mobility of nodes and the mobility of code.

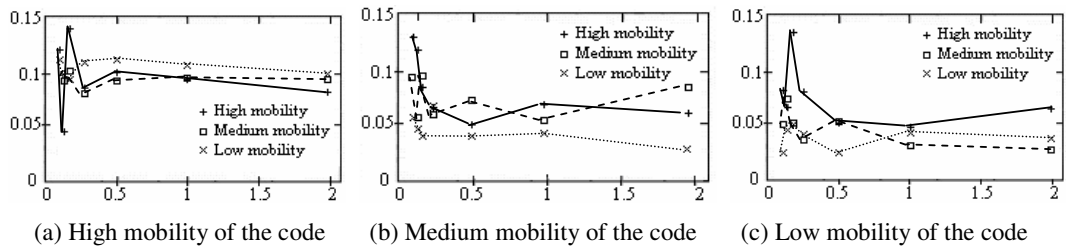

|     |     |     |
| --- | --- | --- |
| (a) High mobility of the code | (b)  Medium mobility of the code | (c)  Low mobility of the code |

Figure 6.   Response time of localization according to the requests load (reactive protocol)

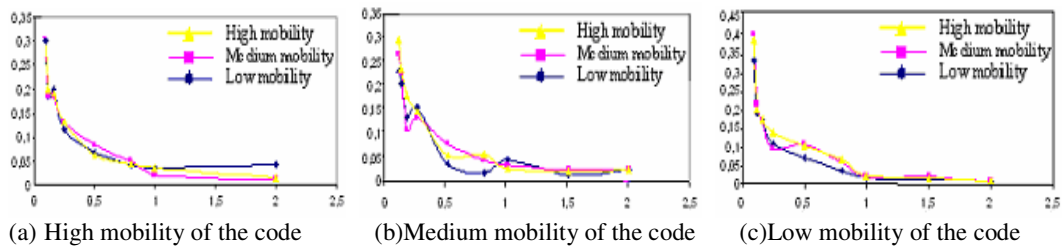

|     |     |     |
| --- | --- | --- |
| (a) High mobility of the code | (b)Medium mobility of the code | (c)Low mobility of the code |

Figure 7.   Response time of localization according to the requests load (centralized protocol)

We remark also that the response time is better for the distributed protocol than for the centralized one. One part of the great number of exchanged messages in the centralized approach case concerns server region, this will provoke its overloading (bottleneck); so, this reduces the server performances, in term of response time, in the processing of localization requests.

- **Discussion**

The comparative study between the two protocols shows that the distributed protocol outperforms the centralized protocol, especially in term of the messages number.

The proactively update of the databases localizations at the level server, generates a huge number of control and diffusion message, and this overload the server. Consequently, the performances of the centralized protocol in term of response time are not improved, in spite of server availability, which is supposed satisfy in a better delay the localization requests than in the distributed protocol case.

After analyzing the causes of these disperformances, we proposed the distribution of the centralized protocol as described in what follow.

## 5. DISTRIBUTED FORM OF THE CENTRALIZED PROTOCOL

This consists of a multi-servers management of the localization problem by reparting the network into many zones. The aim is to assign each server to localize mobile codes of some geographical part of the network (a zone). For this purpose, each server will hold a subset of database containing the mobile codes positions of its zone (figure 8).

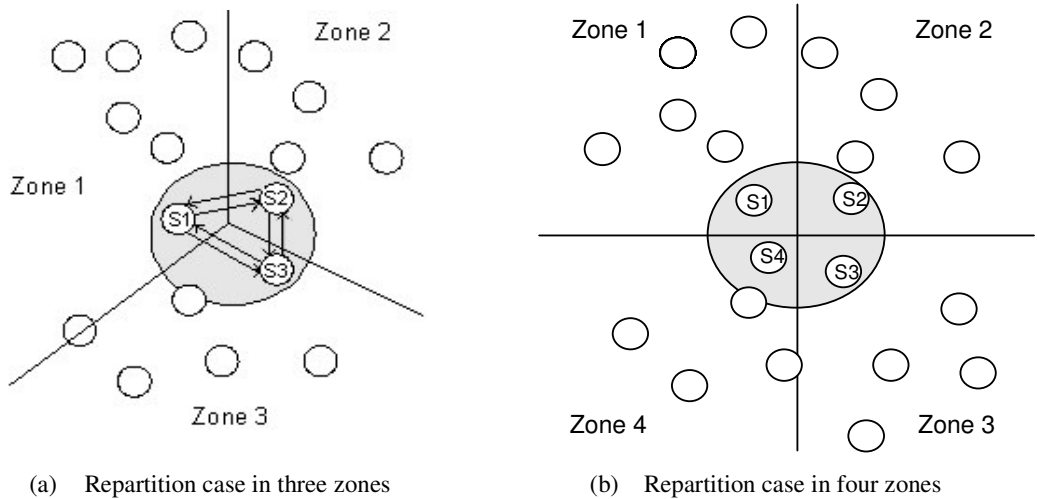

(a)   Repartition case in three zones          (b)   Repartition case in four zones

Figure 8.   Repartition of the network into zones

The principal idea of this distribution is to limit the scope of the diffusions during the localization process to the sole involved zone, thereby avoiding the overload of the localization server. The steps of the protocol are as follows:

*1- Determination of gravity centre of network*
The calculation of gravity centre of the network is in the same manner as in the case with a single server (see paragraph 3.2). Here, the centre of gravity will serve as a benchmark for defining the different areas and for the election of servers associated with each zone.

*2- Describing a zone*
Each zone is represented by a geographical area bounded by two terminals B1 and B2, with as common axis the gravity centre of the network and forming an angle $\alpha$, where:  $0 < \alpha <= \pi$
Assuming that all areas have the same geographical extent, it will $\alpha = 2\pi / n$, $n$ is the number of zones in the network with $n > = 2$.

*3- Assignment nodes of the network to zones*
Are two angles $\beta1$ and $\beta2$ associated respectively to the two terminals B1 and B2 of a zone (see figure 9), such as:                    $\alpha = \beta2 - \beta1$, with  $\beta2 > \beta1$
We will say that a node $i$ belongs to a given zone, bounded by B1 and B2, if its position $(X_i, Y_i)$ check the following two equations:

$$(Y_i - Y_v) \, cos(\beta1) > (X_i - X_v) \, sin(\beta1)$$
$$(Y_i - Y_v) \, cos(\beta2) < (X_i - X_v) \, sin(\beta2)$$

Where $(X_v, Y_v)$ are the coordinates of the centre of gravity of the network.

### 4- Election of heads of zone (servers) and cooperation

In each zone of the network, elect a node closest to the centre of gravity as a server of this zone (the leader of the group). The calculation of the distance to a node from the centre of gravity is in the same manner as in 3.2.

Various selection criteria can be selected for the election of a server, for example, elect a node positioned in the centre of the zone, facilitating communication of the server with all the nodes in the zone. However, cooperation between servers becomes slower than if they were closer to one another, for example by forming a virtual ring as shown in figure 8.b. The cooperation between the different servers is needed to localize mobile codes outside their zone, as well as the setup of database localizations during the change of moving code zones or of the host nodes.

### 5- Using mobile agent and migration

A mobile agent is placed on each node elected as a server of a zone witch will have as a task the management of a sub database (SDB) containing the positions of mobile code inside its zone (i.e. {code_identity, host node_identity}).

When the mobile node is away from the centre of the network or for energy loss battery causes, another node is elected for its replacement. Then, the agent placed on the replaced station migrates towards the elected station, where it continues the sub database management.

### 6- Updating the position's database

The update of the SDB is at each mobility of the code:

-When the code moves to a node in the same zone, its position (identity of the home node) is updated in the SDB server of that zone.

-If the home node is located in another zone, then its position will be inserted in the SDB server destination zone, and deleted from the server zone it left.

-When a node hosting a code change zone, the SDB will be updated in the same manner as if the code has moved to a node in another zone.

### 7- Localization process

The mobile code localization will then take place by sending a request to the server in the same zone which by consulting its SDB, gives the identity of the node hosting the mobile code. If it can not find that information, it forwards the request to a server neighbour, depending on the configuration of the virtual ring, which processes the request in the same way.

Once the identity of the node hosting the code is known, the localization of the node in the network is made with the localization service GLS, perfectly suited to our network geographically divided into zones (grids).

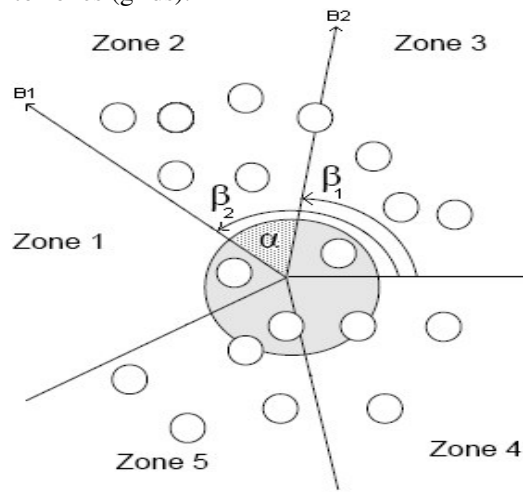

Figure 9.  Description of geographic zone

One important point in this protocol is that a node in its zone has not to know the identity of its zone's server when it sends him a message or a request. Since the request contains the identity of zone, it will concern only the server of this zone.

Note that this protocol assume that there is no partitioning in the ad hoc network; i.e. all the nodes of the network belong to a same and unique partition.

### 5.3 Experimentation

The available performance results in this paper, is only concerned with the case of network repartition in two zones. Other works (in progress) will address the case of repartition in to more than two zones.

The figures 10 and 11shows the performance of the protocol respectively in terms of average messages number and response time per localization request, according to the requests load, the mobility of nodes and the mobility of code.

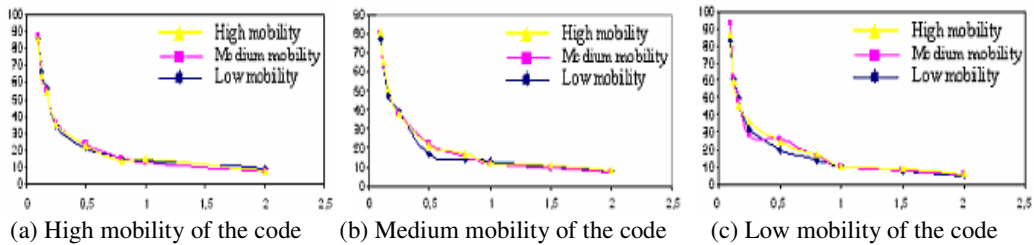

(a) High mobility of the code    (b) Medium mobility of the code    (c) Low mobility of the code

Figure 10.   Average message number according to requests load

Unlike to centralized protocol with one partition, an important improvement in term of number of messages is obtained with our protocol (one for ten times). However, we have not reached the performances of the distributed protocol. As concerns the response time, the results obtained are more encourageant, as they are similar as those obtained with distributed protocol and better than those reported for the centralized protocol with one partition (one for two times).

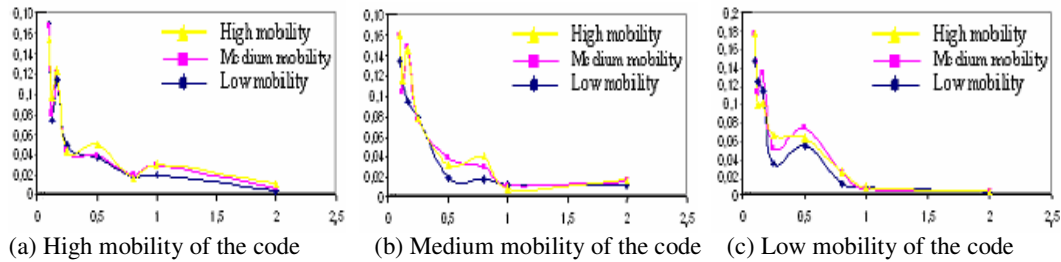

(a) High mobility of the code    (b) Medium mobility of the code    (c) Low mobility of the code

Figure 11.   Response time of localization according to the requests load

## 6. CONCLUSION AND FUTURE WORK

In this paper, we proposed a new approach in the management of mobile ad hoc networks. Our alternative, based on mobile agent technology, allows the design of mobile centralized server in ad hoc network, where it is not obvious to think about a centralized management. The comparative study, between the centralized and the distributed localization protocols in term of messages number exchanged and response time has shown that the centralized approach in a distributed form is more interesting than a totally centralized approach.

Although the centralized approach is yet costly with its current version (two servers), the aim of this approach is to provide permanent availability of services in ad hoc networks.

Having dealt with the distributed of the centralized protocol of localization of mobile codes in the case of two partitions, we are studying the performance of the protocol considering more than two zones.

**Authors**
Dr. Youcef Zafoune

I am an associate professor of computer science at the University of Algiers.
My research focuses on the impact of mobile agent technology on the management
of the mobile ad hoc network.

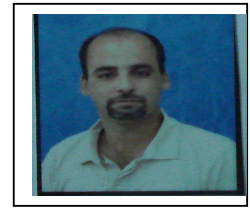

Pr. Aicha Mokhtari is professor of computer science at the University of Algiers.

Dr. Rushed kanawati is associate professor of computer science at the University of Paris13.